\documentclass[5p,10pt]{elsarticle}

\journal{Technical Report}
\usepackage[T1]{fontenc}
\usepackage{amsmath}
\usepackage{amsfonts}
\usepackage[algo2e,ruled,lined]{algorithm2e} 
\usepackage{booktabs}
\usepackage{multirow}
\usepackage{adjustbox}
\usepackage{ulem}
\usepackage{hyperref}
\usepackage{color}

\usepackage{graphicx}
\graphicspath{{./}{./figs/}}

\newcommand{\sstitle}[1]{\smallskip\noindent\textbf{#1.\/}}

\begin{document}

\begin{frontmatter}

\title{Hypergraph Diffusion for High-Order Recommender Systems}
\author{Darnbi Sakong}
\author{Thanh Trung Huynh}
\author{Jun Jo}

\begin{abstract}
Recommender systems rely on Collaborative Filtering (CF) to predict user preferences by leveraging patterns in historical user-item interactions. While traditional CF methods primarily focus on learning compact vector embeddings for users and items, graph neural network (GNN)-based approaches have emerged as a powerful alternative, utilizing the structure of user-item interaction graphs to enhance recommendation accuracy. However, existing GNN-based models, such as LightGCN and UltraGCN, often struggle with two major limitations: an inability to fully account for heterophilic interactions, where users engage with diverse item categories, and the over-smoothing problem in multi-layer GNNs, which hinders their ability to model complex, high-order relationships. To address these gaps, we introduce WaveHDNN, an innovative wavelet-enhanced hypergraph diffusion framework. WaveHDNN integrates a Heterophily-aware Collaborative Encoder, designed to capture user-item interactions across diverse categories, with a Multi-scale Group-wise Structure Encoder, which leverages wavelet transforms to effectively model localized graph structures. Additionally, cross-view contrastive learning is employed to maintain robust and consistent representations. Experiments on benchmark datasets validate the efficacy of WaveHDNN, demonstrating its superior ability to capture both heterophilic and localized structural information, leading to improved recommendation performance.
\end{abstract}
\end{frontmatter}
\section{Introduction}

Collaborative Filtering (CF) is a task of identifying or ranking unobserved items that users would like to engage in based on the behaviors and preference records of similar users. Various applications such as social network services~\cite{huang2021knowledge}, streaming platform~\cite{liu2021concept}, and retail websites~\cite{huang2019online} adopt CF to support efficient user decision processes over the excessive quantity of information. 
By scoring preferences on items based on learning the low-dimensional vector representations of users and items based on the user-item interaction history, conventional CF techniques~\cite{ma2008sorec} have gained significant attention from the community due to their effectiveness.

With the emergence of graph neural networks(GNNs), recent research has started to project representations by harnessing the topological structures inherent in user-item interaction data to capture graph-based collaborative signals. For instance, LightGCN~\cite{he2020lightgcn} and UltraGCN~\cite{mao2021ultragcn} introduce simplified graph convolutional networks(GCNs) to tailor general GCNs to recommendation tasks. 
To further enrich the expressiveness of embeddings, hypergraphs have been introduced in the CF paradigm by its capability to model group-wise relationships between users and items. Unlike ordinary binary graphs, any number of users that have shown their interest in a specific item $i$ are connected with corresponding hyperedge reflecting their multi-way interactions. One of the representative hypergraph-based CF methods, HCCF~\cite{xia2022hypergraph}, proposed a self-supervised learning approach to contrast between local and global embeddings to enhance the generalization ability of the model.

Despite the advancement of GNN-based methods, existing literature often overlooks heterophilic patterns which are often observed in user-item interaction data. In most practical scenarios, items that one specific user has engaged with are in different categories. To estimate the similarity between item representations, embeddings of items in similar or identical categories are expected to be in proximity, while those in different categories are expected to be far apart. Moreover, capturing high-order multi-hop neighborhood structures is important for learning accurate representations. However,  widely adopted stacked multi-layers of HGCN in the existing models often suffer from smoothing problems where embeddings become indistinguishable. To mitigate this, a more nuanced approach is required to naturally balance local and global relationships without stacking layers while considering heterogeneity patterns in interaction graphs~\cite{nguyen2019maximal,nguyen2021judo,nguyen2020factcatch,nguyen2022survey,nguyen2024manipulating}.

To overcome the aforementioned challenges, we introduce a wavelet-based hypergraph diffusion framework for modeling localized structure-aware heterophilic user-item group-wise dependency. More specifically, we fuse two separated channels: \textit{Heterophily-aware Collaborative Encoder} for heterophilic pattern modeling and \textit{Multi-scale Group-wise Structure Encoder} for localized structure modeling.
In summary, the key contributions of this work are organized as below:
\begin{itemize}
    \item In this work, we propose a wavelet-based hypergraph diffusion model called WaveHDNN to capture both heterophilic patterns and localized topological information via simultaneous learning on two separated encoders.

    \item We present a Heterophily-aware Collaborative Encoder that harnesses an equivariant operator inspired by ED-HNN~\cite{wang2022equivariant} to differentiate messages passed to heterogeneous nodes.

    \item We introduce a Multi-scale Group-wise Structure Encoder that tunes the spread of information across the hypergraph more flexibly via wavelet transform~\cite{sun2021heterogeneous} combined with hypergraph convolutional layer for robust structure information learning.

    \item To unify the varied features captured from two encoders, we apply cross-view contrastive learning to ensure the consistency of embeddings of the same entities.

    \item We conducted extensive experiments with state-of-the-art baseline models on three popular recommendation datasets to verify the superior performance of the proposed model.
\end{itemize}

The remainder of the paper is organized as follows. \autoref{sec:related} discusses related work. \autoref{sec:method} presents the overview and specific wavelet-based diffusion framework that captures high-order relationships. \autoref{sec:evaluations} provides empirical evaluation results. Finally, \autoref{sec:conclusion} summarizes and concludes the paper.

\section{Related Works}
\label{sec:related}

\sstitle{Conventional Collaborative Filtering Methods}
Collaborative filtering (CF) is a popular technique in recommender systems, predicting user preferences based on historical user-item interactions~\cite{nguyen2015result,nguyen2017retaining,tam2019anomaly,nguyen2020monitoring,nguyen2019user,nguyen2020entity,nguyen2021structural}. CF methods are categorized into memory-based and model-based approaches. Memory-based methods, like User-Based and Item-Based CF, use similarity measures such as cosine similarity or Pearson correlation~\cite{breese2013empirical,sarwar2001item}. While simple and intuitive, they struggle with scalability and data sparsity in large datasets~\cite{shi2014collaborative}.
Model-based methods, such as Matrix Factorization (MF), project users and items into a shared latent space, offering scalable solutions. BiasMF~\cite{koren2009matrix} adds bias terms for users and items, improving on standard MF by capturing individual characteristics. Singular Value Decomposition (SVD)~\cite{koren2009matrix} decomposes the user-item interaction matrix into latent factors, and SVD++~\cite{koren2008factorization} incorporates implicit feedback like clicks or views. Non-Negative Matrix Factorization~\cite{zhang2006learning} imposes non-negativity constraints, ensuring interpretable, positive representations.
 To capture non-linear user-item relationships, Neural Collaborative Filtering (NCF)~\cite{he2017neural} replaces MF's inner product with a neural network, allowing for complex interactions through multi-layer transformations. However, MF and NCF treat user preferences as static, which limits their effectiveness in dynamic environments. Sequence-based models like Recurrent Neural Networks~\cite{hidasi2015session} and Transformer-based models~\cite{kang2018self} address this by modeling the sequential nature of user interactions, accounting for evolving preferences. Despite the rise of complex models, BiasMF and NCF remain foundational techniques in CF, balancing performance and interpretability while often serving as the basis for advanced hybrid models~\cite{thang2022nature,duong2022efficient,nguyen2020factcatch,hung2017answer,nguyen2017argument}.

\sstitle{GNN-based Collaborative Filtering Methods}
Graph structures are widely used to capture user preferences, representing users and interacted items as nodes connected by edges~\cite{huynh2021network,duong2022deep,nguyen2023example,trung2022learning}. Conventional methods like ItemRank~\cite{gori2007itemrank} assign weights based on random walks, while SimRank~\cite{jeh2002simrank} computes item-user similarity using common neighbors. More recently, Graph Neural Networks (GNNs) have enhanced Collaborative Filtering (CF) by effectively capturing user-item interactions.
NGCF~\cite{wang2019neural} extends GCNs by deepening the propagation process in the user-item graph, and AGCN~\cite{feng2019attention} uses attention mechanisms to emphasize node relatedness. LightGCN~\cite{he2020lightgcn} simplifies GCNs by focusing on embedding propagation, while UltraGCN~\cite{mao2021ultragcn} further simplifies the model for efficiency. PinSage~\cite{ying2018graph} combines random walks and graph convolutions for large-scale recommendation systems.
To address common GNN challenges like over-smoothing, GraphRec~\cite{fan2019graph} introduces attention to focus on important neighbors, and Star-GCN~\cite{zhang2019star} uses a graph auto-encoder for unseen users and items. GCCF~\cite{chen2020revisiting} incorporates residual networks to mitigate over-smoothing. Despite these advancements, existing models often overlook heterogeneity and higher-order connectivity in user-item graphs. Our proposed model addresses these challenges by using a novel message-passing algorithm to capture high-order relationships~\cite{zhao2021eires,huynh2021network,duong2022deep,nguyen2022model,nguyen2022detecting,trung2022learning,huynh2023efficient}.

\sstitle{HGCN-based Collaborative Filtering Methods}
Beyond binary user-item connections, group-level relationships often provide valuable high-order information. A hypergraph, where multiple nodes are connected via hyperedges, captures such relationships and has proven effective in offering rich structural insights. Hypergraph Convolutional Networks (HGCNs)~\cite{feng2019hypergraph} extend Graph Neural Networks (GNNs) for Collaborative Filtering (CF), addressing the limitations of traditional graphs by capturing more complex relationships.
Building on HGCNs, DHCF~\cite{ji2020dual} employs a divide-and-conquer strategy to learn user and item embeddings simultaneously, while HCCF~\cite{xia2022hypergraph} introduces self-supervised learning with contrastive learning to align global and local embeddings. SHT~\cite{sht2022} leverages a hypergraph transformer to incorporate multi-scale neighborhood features, while HypAR~\cite{jendal2024hypergraphs} provides explainable recommendations using HGCN layers. UPRTH~\cite{yang2024unified} utilizes task-specific hypergraphs with transitional attention, and MHCN~\cite{yu2021self} enhances social recommendation by capturing high-order user relations via a multi-channel HGCN.
Despite these advances, further improvements are possible. Our model propagates permutation equivariant messages through HGCN layers, making embeddings more distinguishable. Additionally, wavelet-transform-inspired HGCN layers capture localized topological information across varying scales, further enhancing the model's performance~\cite{nguyen2023poisoning,nguyen2023example,nguyen2014reconciling,nguyen2015smart,thang2015evaluation,nguyen2015tag,hung2019handling}.

\section{Methodology}
\label{sec:method}
\begin{figure}[!h]
	\centering
	\includegraphics[width=1.05\linewidth]{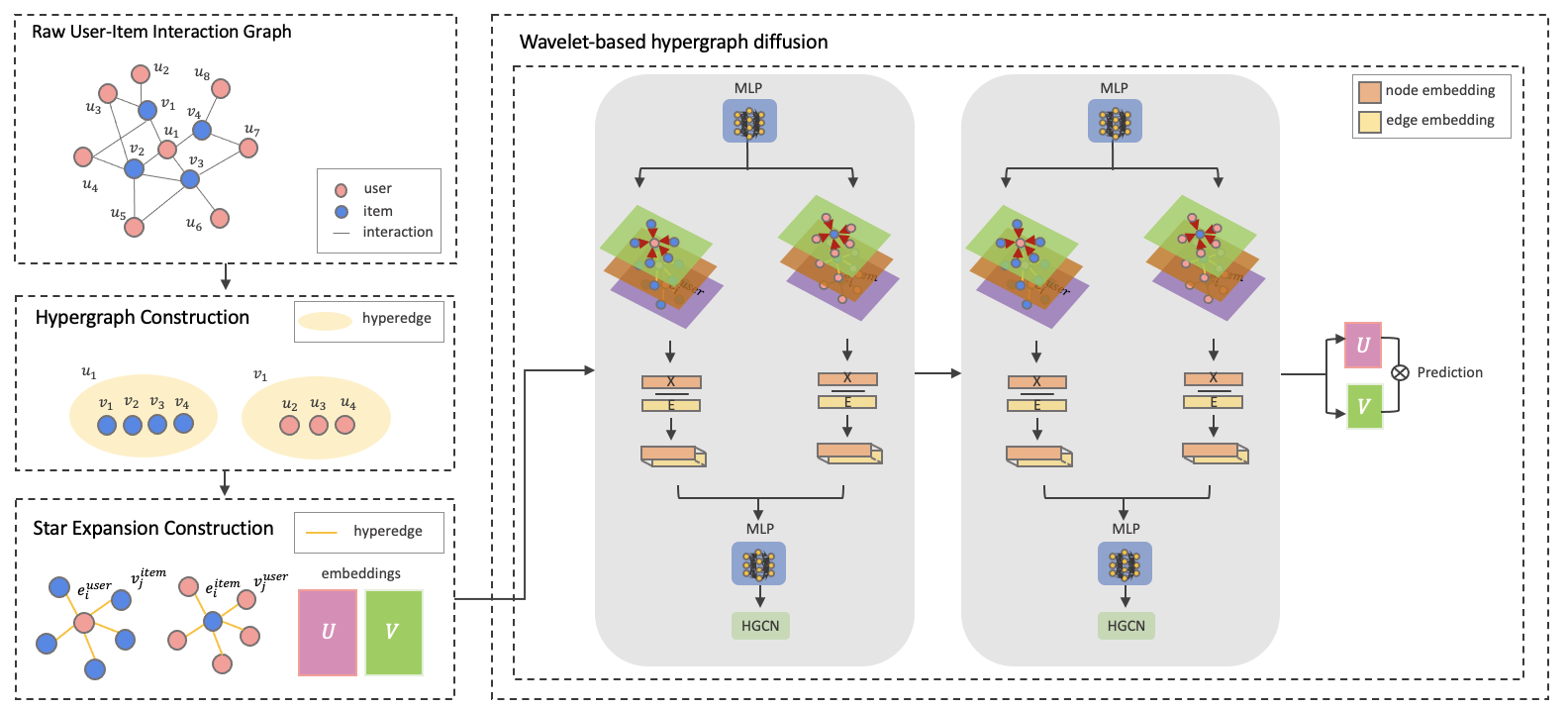}
	\caption{An illustration of WaveHDNN.}
	\label{fig:framework}
\end{figure}

In this section, we introduce our proposed end-to-end framework, namely WaveHDNN, which consists of two individual channels to learn high-order dependencies. \autoref{fig:framework} demonstrates the overall pipeline of WaveHDNN. First, we capture the heterophilic pattern inherent in collaborative hypergraphs based on the heterophily-aware hypergraph diffusion channel. Second, we leverage the channel based on the wavelet-based hypergraph neural networks to capture the user-item group structural relationship at different local and global scales. We further ensure consistency between the two-faceted representations of users and items through contrastive learning and integrate the final representations to predict user preferences.

\subsection{Heterophily-aware Collaborative Encoding}
Building on the principles of ED-HNN~\cite{wang2022equivariant}, we begin by learning user and item representations through the modeling of heterophilic patterns within collaborative hypergraphs. It is important to note that different types or categories of items are often grouped by specific users, a phenomenon commonly observed in user-item collaborative hypergraphs. To differentiate between items while preserving the common features within the same categories, we propose a novel approach that captures heterogeneity by leveraging \textit{Heterophily-aware Hypergraph Diffusion Networks layers}. First, we transform the embeddings of users (or items) leveraging a multi-layer perceptron (MLP). By applying MLP before gathering the node features, we enable the model to adapt to complex patterns in data rather than relying on fixed or predefined parameters, which may not generalize well to non-homophilic scenarios. Then the transformed embeddings are fed into the hypergraph convolution layer(HGCN) to aggregate node features. The learned node features are then concatenated with the initial node embeddings to guarantee an equivariant operator so as to pass varying messages to nodes in hyperedge. Mathematically, the HDNN framework works as:

\begin{align}
\mathbf{X_{e}^{(l)}} = \operatorname{LN}(\operatorname{HConv}(\operatorname{MLP}_1(\mathbf{X^{(l)}}), \mathbf{H})) + \mathbf{X^{(l)}}, \\
\mathbf{X_{v}^{(l)}} = \operatorname{LN}(\operatorname{HConv}(\operatorname{MLP}_2(\mathbf{X_{e}^{(l)}}), \mathbf{H})) + \mathbf{X_{e}^{(l)}}
\end{align}
where $X$ is embeddings of user and items, $X_{e}$ denotes hyperedge embeddings, $X_{v}$ denotes node embeddings, $H$ is incidence matrix, $MLP_{1}(\cdot), MLP_{2}(\cdot)$ indicates MLPs, $HConv(\cdot)$ denotes hypergraph convolution layer, and $LN(\cdot)$ represents Layer Normalization~\cite{xu2019understanding}. 
The learned node embeddings are then further integrated with their original features to prevent information dilution. Finally, we apply the final MLP to update embeddings.

\subsection{Multi-scale Group-wise Structure Encoding}

Following the localized hypergraph neural network proposed in \cite{sun2021heterogeneous}, we propose a wavelet-based multi-scale group-wise relationship aware encoder to capture varied scales of neighborhood structure while preserving heterogeneity in hypergraphs.
Wavelet basis enables localized convolution on the vertex domain, meaning that the convolution operation can focus on specific regions of the hypergraph rather than considering the entire graph structure. This is particularly important for hypergraphs where different types of hyperedges exist, and a localized approach can effectively capture the heterogeneity of the data while encapsulating the topological information into embeddings. The wavelet-based hypergraph convolution layer can be expressed as below.

\begin{align}
    \mathbf{X}^{(l+1)} = \mathbf{\Theta\Lambda\Theta^{\prime}X^{(l)}W} + \mathbf{X}^{(l)}
\end{align}
where $\mathbf{X}$ denotes feature, $\mathbf{\Lambda}$ denotes diagonal weight filter matrix, $\mathbf{\Theta}$ and $\mathbf{\Theta^{\prime}}$ denote wavelet and its inversed version, and $\mathbf{W}$ denotes weight matrix for feature transformation.
Unlike wavelet HGCN in \cite{sun2021heterogeneous}, we concatenate the current embedding of nodes with its newly learned representations in the layer to avoid signal degradation through multiple layers.

\subsection{Optimization}
The two separated channels capture different aspects of user behaviors and keeping the embeddings of the identical users and items learned from each encoder close ensures that diverse perspectives reflect a unified representation. This allows the model to form a consistent and coherent comprehension of user preferences. Besides, if the embeddings from different modules are far apart, it may indicate the features captured by each module are redundant or even conflicting information. To preserve the unique contribution of each encoder, embeddings need to be close in proximity within the vector space. Hence, we adopt a cross-view contrastive learning mechanism under the assumption that the embeddings of corresponding users(or items) from different encoders as positive pairs while those that are not as corrupted negative pairs. The contrastive loss inspired by InfoNCE~\cite{InfoNCE} can be represented below.

\begin{equation}
    \label{eqn:ssl_u}
    \mathcal{L}_s^{(u)}=\sum_{i=0}^I \sum_{l=0}^L-\log \frac{\exp \left(s\left(\mathbf{z}_{i, l}^{(u)}, \Gamma_{i, l}^{(u)}\right) / \tau\right)}{\sum_{i^{\prime}=0}^I \exp \left(s\left(\mathbf{z}_{i, l}^{(u)}, \Gamma_{i^{\prime}, l}^{(u)}\right) / \tau\right)}
\end{equation}

where $\tau$ refers to a temperature parameter, $s(\cdot)$ represents the cosine similarity function, $\mathbf{z}{i,l}$ denotes the collaborative latent vector of user(or item) at layer $l$, and $\Gamma{i,l}$ represents the latent vector of the same user(or item) from each encoder at the same layer.

We further opt for widely-used BPR loss~\cite{BPR} which assumes that if user $u$ has interacted with item $i$, then the user prefers $i$ over other items that the user has not interacted with before. To learn the model parameter, we minimized the BPR loss function expressed as below.

\begin{equation}
    \mathcal{L}_{\text{BPR}} = \sum_{u \in \mathcal{U}} \sum_{i \in \mathcal{I}_u} \sum_{i' \notin \mathcal{I}_u} - \log \sigma(\hat{y}_{u,i} - \hat{y}_{u,i'})
\end{equation}
where $\mathcal{U}$ denotes a set of users, $i \in \mathcal{I}_u$ denotes interacted items while $i' \notin \mathcal{I}_u$ denotes items that interaction with user has not been observed. $\hat{y}$ is the probability that user $u$ might be inclined to engage with item $i$, which is computed by the dot product between user and item embeddings.
\section{Evaluation}
\label{sec:evaluations}

In this section, we evaluate the effectiveness of WaveHDNN against selected state-of-the-art baselines. We first describe the experimental setting and then present empirical performance comparison on different datasets. To rationalize the design of the architecture of the model, we conduct an ablation test by replacing each key component of the model.

\subsection{Experimental Setting}
For the fairness of the comparison, we perform evaluations on three different real-world datasets: Amazon-books for book recommendations, Steam for game recommendations, and Yelp for business recommendations. The comprehensive statistics of datasets are presented in \autoref{tab:datasets1} The varying levels of density in real-world interactions within the selected datasets highlight the robustness of our model across different practical conditions. We divided each dataset into training, validation, and test sets following the 7:1:2 ratio and used an average of 5 times running as the final performance of each model.

\begin{table}[!h]
\vspace{-1em}
\centering
\footnotesize
\caption{Statistics of Recommendation Datasets}
\label{tab:datasets1}
\begin{tabular}{c | c c c}
    \toprule
    Statistics & Amazon-Books & Steam & Yelp\\ 
    \hline 
    \#Users & 11,000 & 23,310 & 11,091 \\ 
    \#Items  & 9,332 & 5,237 & 11,010\\ 
    \#Interactions & 200,860 & 525,922 & 277,535\\ 
    Density & $1.9 \times 10^{-3}$ & $4.30 \times 10^{-3}$ & $2.2 \times 10^{-3}$\\ 
    \bottomrule
\end{tabular}
\vspace{-1em}
\end{table}

\subsection{Base Models}
We assess the effectiveness of our WaveHDNN with 5 state-of-the-art baselines that cover GNN-based and Hypergraph-based Recommendation tasks.

\begin{itemize}
    \item \textbf{LightGCN} eliminates the redundant components by preserving only the neighborhood message passing component from NGCF for CF-oriented representation learning.
    \item \textbf{SGL} enhances the LighGCN framework with an augmentation mechanism and self-supervised contrastive learning.
    \item \textbf{DHCF} introduces a divided-and-conquer method with dual-channel hypergraph neural networks to enable disjoined and simultaneous learning of user and item embeddings.
    \item \textbf{HCCF} encodes local and global views of user-item hypergraphs with contrastive learning to learn distinguishable user and item representations.
    \item \textbf{SHT} adopts transformer mechanism and applies hypergraph attention strategy. The model further presents a data augmentation approach by complementing different scopes of view of CF signals.
    \item \textbf{AutoCF} designs automatic data augmentation with generative self-supervised learning to automatically extract significant self-supervised features.
\end{itemize}

\subsection{Performance Comparison}

This section provides a comprehensive analysis of the performance results of all baseline models on three datasets and summarize the comparison between them.

\sstitle{Overall Comparison} 
As shown in \autoref{tab:baseline_evaluation1}, \autoref{tab:baseline_evaluation2}, and \autoref{tab:baseline_evaluation3}, our proposed WaveHDNN model consistently outperforms all baselines across six evaluation metrics, demonstrating its robustness in various scenarios. Notably, WaveHDNN surpasses the second-best model by $7.24\%$ on the Steam dataset in terms of NDCG@20, a key metric for ranking item relevance within the top 20 recommendations. 
In sparse datasets like Amazon-books and Yelp, with fewer user-item interactions, WaveHDNN shows significant improvements of $3.92\%$ and $5.45\%$ over the second-best models. These gains emphasize the model's adaptability, even in challenging environments.

WaveHDNN's superior performance can be attributed to key factors: 1) Learning distinguishable representations through equivariant operator-based hypergraph convolution networks while preserving local and global neighborhood messages. 2) Incorporating different scales of localized features via a combination of hypergraph convolution and wavelet-based hypergraph transform layers for more expressive feature learning. These advantages highlight WaveHDNN's versatility and its ability to adapt to varying interaction sparsity and dataset characteristics, making it highly effective for collaborative filtering tasks.

\sstitle{Superiority over Hypergraph-based baselines}
The experimental results clearly show that hypergraph-based methods outperform traditional graph-based models, such as LightGCN and SGL, in all test cases. This gap highlights the advantage of using hypergraphs to capture more complex user-item interactions compared to conventional binary graphs. Hypergraphs model high-order group interactions, which standard graph methods often miss, leading to more accurate recommendations.

In particular, SHT, a hypergraph-based method, surpasses SGL by $34.7\%$ and $21.2\%$ on NDCG@40 for the Amazon-books and Steam datasets, respectively. This improvement underscores hypergraphs' ability to capture intricate group relationships and higher-order dependencies, resulting in more precise recommendations. Among hypergraph-based methods, WaveHDNN stands out, achieving an $8.6\%$ improvement on Recall@20 on the Amazon-books dataset, demonstrating its superior ability to retrieve relevant items for users. Recall@20 reflects the model's effectiveness in recommending items users are likely to engage with. This improvement is due to WaveHDNN's advanced integration of equivariant message passing and wavelet-based structure learning.

\begin{table*}[!h]
\vspace{-1em}
\centering
\footnotesize
\caption{The overall performance comparison results for WaveHDNN and compared baseline models on the Amazon-books dataset. The top mark performance is highlighted with bold text and runner-up performances are underlined with borderline.}
\label{tab:baseline_evaluation1}
\begin{adjustbox}{max width=1.0\linewidth}
\begin{tabular}
{@{}ccccccc@{}ccccccc@{}ccc@{}}\midrule 
\multirow{2}{*}{Model}  & \multicolumn{6}{c}{Amazon-books}\\
\cmidrule{2-7}
&Recall@10&NDCG@10&Recall@20&NDCG@20&Recall@40&NDCG@40\\ \midrule
\text{LightGCN} & 0.05276 & 0.04472& 0.08412 & 0.05566 & 0.13369 & 0.071 \\
\text{SGL} &0.06331 & 0.05365 & 0.10035 & 0.06664 & 0.15863 & 0.08471 \\
\midrule
\text{DHCF} & 0.06355 & 0.05345 & 0.09913 & 0.06576 & 0.15285 & 0.08254 \\
\text{HCCF} & 0.07332 & 0.06133 & 0.11695 & 0.07653 & 0.17942 & 0.09599\\
\text{SHT} & 0.09123 & 0.07616 & 0.13929 & 0.09304 & 0.20774 & 0.11418 \\
\midrule
\text{AutoCF} & \underline{0.09765} & \underline{0.0839} & \underline{0.14391} & \underline{0.09999} & \underline{0.20815} & \underline{0.12013}\\
\midrule
\textbf{WaveHDNN} &\textbf{0.10132}& \textbf{0.08657} & \textbf{0.15136} & \textbf{0.10389} & \textbf{0.21812} & \textbf{0.12484} \\
\hline
\midrule
\textbf{\%Improve.} & 3.75\%& 3.18\% & 5.17\%& 3.9\%& 4.78\% & 3.92\% \\
\bottomrule
\end{tabular}
\end{adjustbox}
\end{table*}
\begin{table*}[!h]
\centering
\footnotesize
\caption{The overall performance comparison results for WaveHDNN and compared baseline models on the Steam dataset. The top mark performance is highlighted with bold text and runner-up performances are underlined with borderline.}
\label{tab:baseline_evaluation2}
\begin{adjustbox}{max width=1.0\linewidth}
\begin{tabular}
{@{}ccccccc@{}ccccccc@{}ccc@{}}\midrule 
\multirow{2}{*}{Model}  & \multicolumn{6}{c}{Steam}\\
\cmidrule{2-7}
&Recall@10&NDCG@10&Recall@20&NDCG@20&Recall@40&NDCG@40\\ \midrule
\text{LightGCN} & 0.06503 & 0.06021 & 0.10555 & 0.07535 & 0.16418 & 0.09554\\
\text{SGL} & 0.06541 & 0.06067 & 0.10524 & 0.07571 & 0.16385 & 0.09549\\
\midrule
\text{DHCF} & 0.07393 & 0.06773 & 0.11983 & 0.08515 & 0.18867 & 0.10842\\
\text{HCCF} & 0.07756 & 0.0721 & 0.12503 & \underline{0.08953} & 0.19201 & 0.11195\\
\text{SHT} & \underline{0.08006} & 0.07583 & \underline{0.12687} & 0.02647 & \underline{0.19513} & \underline{0.11579}\\
\midrule
\text{AutoCF} & 0.06562 & \underline{0.07632} & 0.10627 & 0.07615 & 0.16475 & 0.09567\\
\midrule
\textbf{WaveHDNN} & \textbf{0.08607}&\textbf{0.07754}&\textbf{0.13525}&\textbf{0.09602} & \textbf{0.20651} & \textbf{0.12019}\\
\hline
\midrule
\textbf{\%Improve.} & 7.5\%& 1.59\%& 6.6\% & 7.24\% & 5.83\%& 3.79\%\\
\bottomrule
\end{tabular}
\end{adjustbox}
\end{table*}

\begin{table*}[!h]
\centering
\footnotesize
\caption{The overall performance comparison results for WaveHDNN and compared baseline models on the Yelp dataset. The top mark performance is highlighted with bold text and runner-up performances are underlined with borderline.}
\label{tab:baseline_evaluation3}
\begin{adjustbox}{max width=1.0\linewidth}
\begin{tabular}
{@{}ccccccc@{}ccccccc@{}ccc@{}}\midrule 
\multirow{2}{*}{Model}  & \multicolumn{6}{c}{Yelp} \\
\cmidrule{2-7}
&Recall@10&NDCG@10&Recall@20&NDCG@20&Recall@40&NDCG@40 \\ \midrule
\text{LightGCN} & 0.05379 & 0.05072& 0.08854 & 0.06302 & 0.14062 & 0.08065\\
\text{SGL} & 0.0543 & 0.05103 & 0.08839 & 0.06308 & 0.14138 & 0.08093\\
\midrule
\text{DHCF} & 0.05302 & 0.04991 & 0.08855 & 0.06255 & 0.144 & 0.08154\\
\text{HCCF} & 0.06265 & 0.05825 & 0.10426 & 0.07323 & 0.16882 & 0.09505\\
\text{SHT} & \underline{0.06955} & \underline{0.06474} & 0.11281 & \underline{0.08021} & 0.1836 & \underline{0.10403}\\
\midrule
\text{AutoCF} & 0.06694 & 0.06158 & \underline{0.11401} & 0.07856 & \underline{0.18773} & 0.10346\\
\midrule
\textbf{WaveHDNN} &\textbf{0.07393}& \textbf{0.06831} & \textbf{0.1212} & \textbf{0.08502} & \textbf{0.19427} & \textbf{0.10971}\\
\hline
\midrule
\textbf{\%Improve.} & 6.29\%& 5.51\% & 6.3\%& 5.99\%& 3.48\% & 5.45\%\\
\bottomrule
\end{tabular}
\end{adjustbox}
\vspace{-1em}
\end{table*}

\subsection{Ablation Test}
In this section, we evaluate the importance of the main components of our WaveHDNN. To this end, we replace or remove each component to build variation models to showcase how much performance improvement is observed with the proposed components. Results are presented in \autoref{tab:ablation}.

\sstitle{Heterophily-aware Collaborative Encoder}
We investigate the impact of integrating a Heterophily-aware Collaborative Encoder on the performance of learning CF-oriented embeddings for users and items. Specifically, we examine how considering heterophily can enhance the model's representational power in CF. To validate this, we design a variant of the WaveHDNN model without the heterophily-aware encoder, isolating its contribution.
Our results reveal a noticeable performance gap, demonstrating the encoder's effectiveness in learning embeddings within a heterophily-aware framework. The key improvement comes from the diverse messages passed through the model's learnable message-passing algorithm, which operates in the HGCN-based diffusion layers. These layers help prevent over-smoothing by retaining diverse user-item representations, resulting in more accurate predictions and improved performance.
    
\sstitle{Multi-scale Group-wise Structure Encoder} 
    We aim to analyze the role of the Wavelet-based Hypergraph Convolutional Network (HGCN) layer, focusing on its ability to capture multi-scale group-wise structures within a graph. To assess its impact, we build a model variant without the Multi-scale Group-wise Structure Encoder. Experimental results show a significant performance drop when this mechanism is removed, as the model fails to capture complex relationships across neighborhood scales. This degradation underscores the importance of preserving multi-scale structural nuances in the user-item interaction graph.
The observed performance decline is linked to the absence of multi-scale aggregation, which is crucial for retaining important feature information. Without this, features from neighboring nodes are overly smoothed or lose significance through message passing layers. In contrast, the Wavelet-based HGCN layer uses varying scaling parameters to integrate information across different levels of granularity, preserving both local and global structures. This helps capture subtle neighborhood interactions, which improves predictive performance and generalization. The multi-scale aggregation also prevents over-smoothing, enriching the learned representations and enhancing accuracy in collaborative filtering predictions.

\begin{table}[!h]
\vspace{-1em}
\centering
\caption{\textbf{Overview of ablation test of WaveHDNN with its variants.}}
\label{tab:ablation}
\begin{adjustbox}{max width=1.0\linewidth}
\begin{tabular}{@{}lcc@{\hskip 0.5cm}cc@{\hskip 0.5cm}cc@{\hskip 0.5cm}cc@{}}
\toprule
\multirow{2}{*}{Ablation Settings} & \multicolumn{2}{c}{Amazon-books} & \multicolumn{2}{c}{Steam} & \multicolumn{2}{c}{Yelp}\\
\cmidrule{2-3} \cmidrule{4-5} \cmidrule{6-7}
& Recall@40 & NDCG@40 & Recall@40 & NDCG@40 & Recall@40 & NDCG@40\\
\midrule
\text{WaveHDNN} & 0.21845 & 0.12486 & 0.20651 & 0.12019 & 0.19427 & 0.10971\\
\midrule
\text{w/o Heterophily-aware Encoder} & 0.18623 & 0.10627 & 0.19123 & 0.1104  & 0.17236 & 0.09499 & \\
\text{w/o Multi-scale Encoder} & 0.21035 & 0.11626 & 0.18845 & 0.10701 & 0.16856 & 0.09368 \\
\bottomrule
\end{tabular}
\end{adjustbox}
\vspace{-1em}
\end{table}

\subsection{Hyper-parameter Studies}
In this section, we analyze the degree of sensitivity of our model on various hyperparameter settings. mong the standard hyperparameters, we focus on the number of HGCN layers and the embedding dimension, as these parameters significantly impact performance variations.

\begin{figure}[!h]
\vspace{-1.0em}
  \centering
  \begin{minipage}{0.48\linewidth}
    \centering
 	\includegraphics[width=1.03\linewidth]{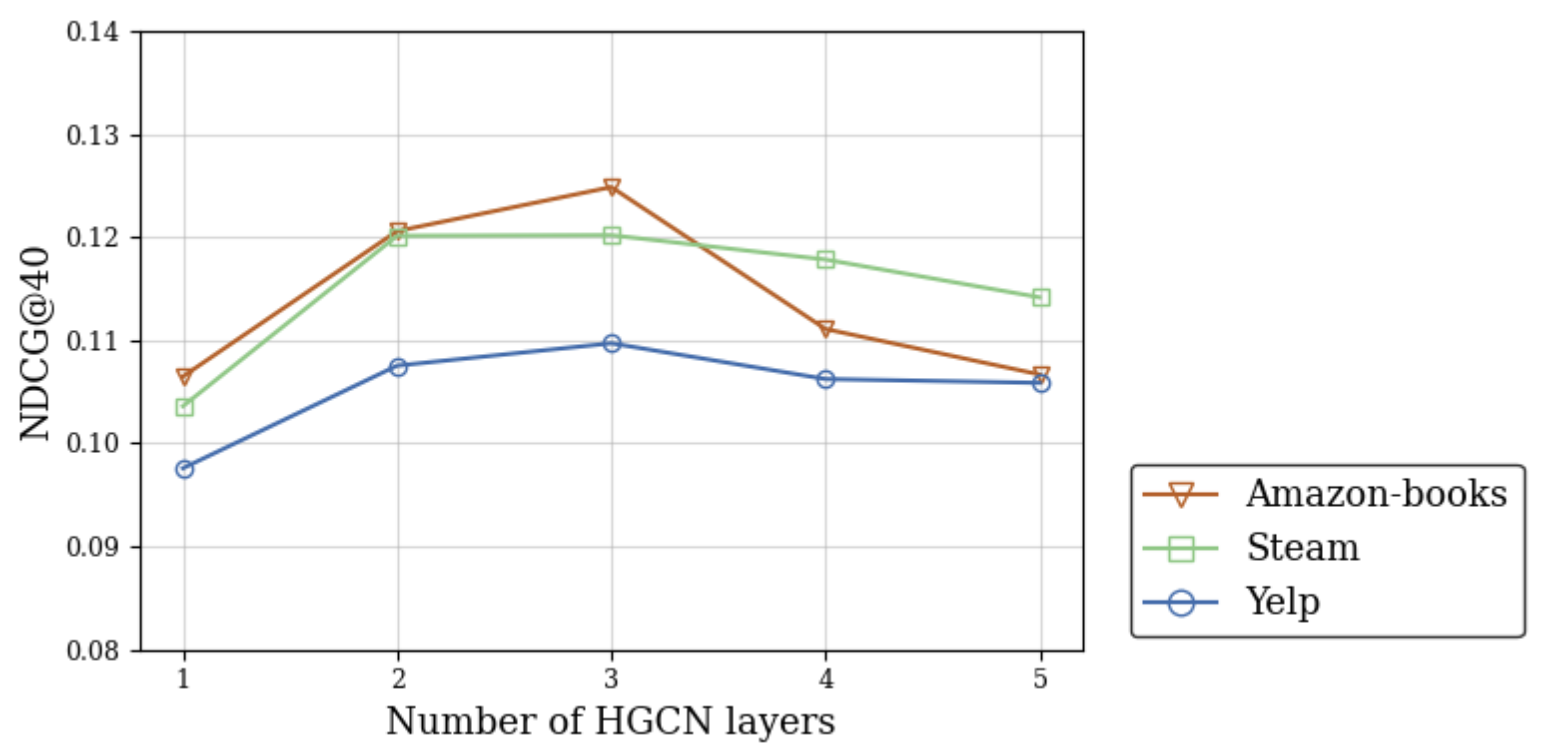}
	\vspace{-1em}
	\caption{Effects of no. HGCN layers.}
	\label{fig:num_layers}
  \end{minipage}
  \begin{minipage}{.48\linewidth}
\centering
\includegraphics[width=1.03\linewidth]{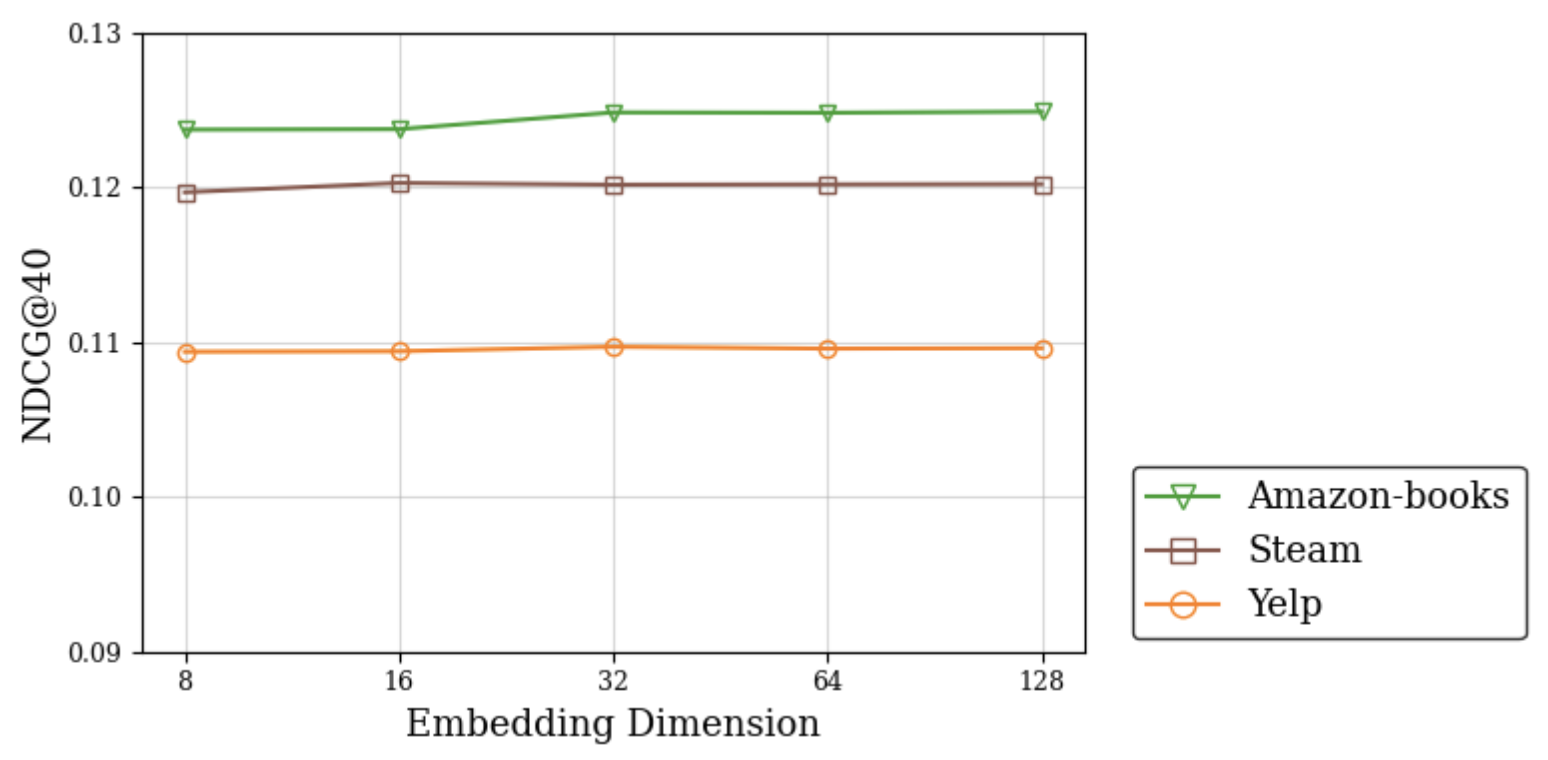}
	\vspace{-1em}
	\caption{Effects of embedding dimensions.}
	\label{fig:emb_size}
  \end{minipage}
\vspace{-1em}
  \end{figure}

\sstitle{Effect of number of HGCN layers.} We run our model with various number of HGCN layers to evaluate the influence of the layer depth. \autoref{fig:num_layers} depicts of performance variations regarding the number of HGCN layers. We can observe that the different number of HGCN layers of each channel has a high impact on recommendation accuracy. We have conducted experiments on three different datasets and the number of layers selected from [1,2,3,4,5]. The results indicate that increasing the number of layers does not necessarily lead to improved performance. The best performance was observed when the number of layers was set to 3, while performance degraded as the layers increased beyond this point. This phenomenon is attributed to overly distinguishable learned representations resulting from the use of a broader neighborhood scope.

\sstitle{Effect of embedding dimensions} We have conducted experiment on three different datsets by changing the embedding dimension from [8,16,64,32,128] to check the influence of different embedding sizes. \autoref{fig:emb_size} shows that varying the embedding size does not always guarantee improved accuracy. Contrary to expectations, the performance improvement was not significant as the embedding size increased. If the embedding size is too large, the important information might get diluted across many dimensions. This can make the model struggle to learn meaningful representations, as the critical data may be spread out excessively thinly.
Besides, as the embedding size becomes excessively large, the memory and computational time requirements increase, leading to inefficient learning.

\section{Conclusion}
\label{sec:conclusion}

In this work, we addressed key limitations in existing Collaborative Filtering (CF) models, particularly those based on Graph Neural Networks (GNNs), which often fail to account for heterophilic patterns and suffer from over-smoothing issues. To tackle these challenges, we proposed a novel wavelet-based hypergraph diffusion model, WaveHDNN, which captures both heterophilic patterns and localized topological information. Our model incorporates two main components: the Heterophily-aware Collaborative Encoder, designed to differentiate message passing for heterogeneous nodes, and the Multi-scale Group-wise Structure Encoder, which leverages wavelet transforms to flexibly tune the spread of information across the hypergraph.
Additionally, we introduced cross-view contrastive learning to ensure the consistency of embeddings across different views. The experimental results on three widely used recommendation datasets demonstrated that WaveHDNN outperforms state-of-the-art CF models by effectively capturing complex user-item interactions and modeling the underlying group-wise dependencies.
In conclusion, WaveHDNN provides a robust and scalable solution for recommendation tasks, offering enhanced generalization and representation capabilities through its innovative use of hypergraph diffusion and multi-scale learning. Future work could explore more complex interaction types~\cite{yang2024pdc,sakong2024higher,huynh2024fast,huynh2025certified,nguyen2023isomorphic,nguyen2024portable} or dynamic environments where user preferences evolve over time~\cite{nguyen2024manipulating,nguyen2025privacy,pham2024dual,nguyen2024multi,nguyen2024handling}.


\appendix

\end{document}